# Magnetoelectric properties of the layered room-temperature antiferromagnets BaMn$_2$P$_2$ and BaMn$_2$As$_2$

S W Lovesey[1,2] and D D Khalyavin[1]

[1]ISIS Facility, STFC Oxfordshire OX11 0QX, UK
[2]Diamond Light Source Ltd, Oxfordshire OX11 0DE, UK

**Abstract** Properties of two ThCr$_2$Si$_2$-type materials are discussed within the context of their established structural and magnetic symmetries. Both materials develop collinear, G-type antiferromagnetic order above room temperature, and magnetic ions occupy acentric sites in centrosymmetric structures (magnetic crystal-class 4'/m'm'm). We refute a previous conjecture that BaMn$_2$As$_2$ is an example of a magnetoelectric material with hexadecapole order by exposing flaws in supporting arguments, principally, an omission of discrete symmetries enforced by the symmetry ($\bar{4}$m'2') of sites used by Mn ions and, also, improper classifications of the primary and secondary order-parameters [H. Watanabe and Y. Yanase, Phys. Rev. B **96**, 064432 (2017)]. Implications for future experiments designed to improve our understanding of BaMn$_2$P$_2$ and BaMn$_2$As$_2$ magnetoelectric properties, using neutron and x-ray diffraction, are examined. Patterns of Bragg spots caused by conventional magnetic dipoles and magnetoelectric (Dirac) multipoles are predicted to be distinct, which raises the intriguing possibility of a unique and comprehensive examination of the magnetoelectric state by diffraction; Dirac multipoles cause Bragg spots with Miller index *l* even (and *h* + *k* even) while magnetic dipoles appear for *l* odd (and *h* + *k* odd). A roto-inversion operation ($\bar{4}$) in Mn site symmetry is ultimately responsible for the distinguishing features.

## I. INTRODUCTION

Structural and magnetic properties of BaMn$_2$P$_2$ [1] and BaMn$_2$As$_2$ [2] are similar to those of the archetypal magnetoelectric material chromium sesquioxide, Cr$_2$O$_3$ [3, 4]. For, the transition metal ions in these materials occupy acentric sites in centrosymmetric structures, and their magnetic dipole moments form simple antiferromagnetic patterns, with magnetic propagation vector **k** = (0, 0, 0), at relatively high temperatures. Moreover, the development of long-range magnetic order as a function of temperature is continuous, without a structural transition or distortion, and hysteresis is absent. While the pattern of dipole moments breaks both time reversal symmetry and inversion symmetry that are present in the paramagnetic phase, the product of the two symmetries is not changed by the onset of magnetic order and the magnetoelectric property is allowed [5, 6, 7]. At an atomic level of detail, Cr ions in Cr$_2$O$_3$ are allowed parity-odd quadrupoles, which can diffract x-rays [8], and the same could be true of Mn ions in BaMn$_2$P$_2$ and BaMn$_2$As$_2$, according to calculations reported for Ba$_{1-x}$Na$_x$Fe$_2$As$_2$ [9].

We offer a discussion of the magnetoelectric properties of BaMn$_2$P$_2$ and BaMn$_2$As$_2$ [1, 2], which present small band gap local moment semiconducting behaviour, that is informed by magnetic symmetry in the footsteps of Dzyaloshinskii [4]. In this context, we recall the

"totalitarian principle" attributed to Murray Gell-Mann by which anything not forbidden (by symmetry) is compulsory, e.g., weak ferromagnetism and Dirac multipoles [10]. The $ThCr_2Si_2$-type chemical structure (I4/mmm, #139) is illustrated in Fig. 1 - it accounts for more than 400 compounds of the $AB_2X_2$ composition - together with the magnetic structure of interest (I4'/m'm'm, #139.536 [11]), while other physical properties of the materials are gathered in Table I. The $ThCr_2Si_2$-type structure is constructed from layers of edge sharing $BX_{4/4}$ tetrahedra alternating with Ba ions. At this point it is of interest to note some significant differences between $BaMn_2P_2$ and $BaMn_2As_2$ and isostructural $BaFe_2As_2$: specifically, $BaFe_2As_2$ presents metallic character and itinerant magnetism, a structural distortion, superconductivity, and moderate Mn-pnictogen hybridization, with approximately 10-20% As character in the d bands.

Referring to Fig. 1, manganese networks in $BaMn_2P_2$ and $BaMn_2As_2$ are square planar. Tetrahedral coordinated Mn ions, with acentric site symmetry is $\bar{4}m2$, form $Mn_2P_2$ and $Mn_2As_2$ sheets separated by the cations. Ba cations sit in an eightfold cubic hole created by phosphorus or arsenide ions. The pattern of magnetic dipoles, aligned along the c-axis, is identical to well-known orthorhombic perovskites. It is G-type antiferromagnetism indexed on the chemical structure, with I4'/m'm'm the appropriate magnetic space-group. Saturation magnetic moments determined from neutron diffraction, listed in Table I, are smaller than the value for high-spin $Mn^{2+}$ ($3d^5$). The observed departure from a value = 5 $\mu_B$ for the moment, derived using $S_z$ = 5/2, can be attributed to covalency combined with antiferromagnetic order, and Hubbard and Marshall discuss various consequences of covalent bonding depending on local crystal symmetry [12].

Our misgivings about theoretical results for magnetoelectric properties of $BaMn_2As_2$ reported by Watanabe and Yanase are on two levels [13]. In the first place, analogy with weak ferromagnetism in haematite ($\alpha$-$Fe_2O_3$) [14], say, leads us to question their assertion that a weak magnetoelectric multipole is a relevant order parameter. The assertion would be justified if magnetic dipoles did not order. This extreme case is realized in the pseudo-gap phase of the ceramic superconductor Hg1201 where conventional, parity-even magnetism using Cu ions is forbidden by anti-inversion, $\bar{1}'$, in the Cu site symmetry, leaving Dirac multipoles as legitimate candidates for an order parameter [15]. No such finding could be reached with the approach to magnetic properties adopted by Watanabe and Yanase [13], where complete omission of the symmetry of the sites used by Mn ions in $BaMn_2As_2$ renders the approach deficient at this level. The omission is a non-trivial shortcoming, because Mn site symmetry ($\bar{4}m'2'$) does not match the magnetic crystal-class (point group) of the space-group (4'/m'm'm).

It is fitting to discuss magnetic properties of materials in terms of a unit-cell structure factor suitable for the interpretation of Bragg diffraction patterns, because well-established scattering techniques yield an abundance of information about magnetic materials. Watanabe and Yanase [13] have calculated magnetoelectric multipoles that are directly relevant to a simulation of neutron diffraction, according to our calculations. The reported non-zero values are grounds for optimism about the success of experiments that we propose. Bulk properties, such as the magnetoelectric effect, are prescribed by the structure factor evaluated for zero deflection of the radiation, i.e., the forward direction. By definition, the unit-cell structure factor incorporates all elements of symmetry in a given magnetic structure, including discrete symmetries demanded by local environments at sites used by magnetic ions.

## II. SYMMETRY INFORMED ANALYSIS

Primary and secondary magnetic order-parameters merit comment. A standard example where both order-parameters are present is weak ferromagnetism allowed by symmetry in a magnetic space-group, and possibly realized in a material by the action of a Dzyaloshinskii-Moriya asymmetric spin interaction. Such symmetry-allowed weak ferromagnetism is secondary to a primary magnetic-order manifest in a simple pattern of magnetic dipoles allowed by symmetry. The secondary order parameter is a coupling phenomenon and does not have its own instability even if the symmetries of the secondary and primary order parameters are identical. For materials under discussion, G-type antiferromagnetism using conventional magnetic dipoles provides the primary order parameter, and magnetoelectric (Dirac) multipoles are secondary, and possibly beyond observation because they are relatively weak. The latter caveat is propelled by the apparent success in refining available neutron diffraction patterns for $BaMn_2P_2$ and $BaMn_2As_2$ against a model that uses magnetic dipoles only [1, 2].

Classifying Dirac multipoles as candidates for a primary order parameter would be justified if magnetic dipoles did not order. This extreme case is realized in the pseudo-gap phase of the ceramic superconductor Hg1201, where conventional, parity-even magnetism using Cu ions is forbidden by anti-inversion, $\bar{1}'$, in the Cu site symmetry [15].

In the present case, Dirac multipoles are shown to contribute to diffraction patterns allowed by the chemical structure. Bragg spots indexed on the chemical structure and observed in the diffraction of neutrons arise from nuclear and magnetic amplitudes, whose contributions can be resolved by polarization analysis. X-ray Bragg diffraction exploiting the intensity enhancement offered by an atomic resonance is possibly a better option for observation of Dirac multipoles. We demonstrate that Dirac multipoles calculated by Watanabe and Yanase [13] are relevant in a feasibility study of diffraction by $BaMn_2As_2$, which gives meaningful ground to optimism about the likely success of experiments that we propose to better understand magnetoelectric properties of $BaMn_2As_2$ and $BaMn_2P_2$.

An analysis of magnetic phenomena informed by symmetry is incomplete and, consequently, potentially misleading if it fails to impose all restrictions, or selection rules, demanded by elements of symmetry in the full magnetic space-group, comprised of rotations, inversions, time reversal and translations. Failure to include discrete symmetries imposed by symmetry in the sites used by magnetic ions renders analysis incomplete in all but one case. The singular case is that of identical site and space group symmetries. For the space group I4/mmm the singular case is realized by ions at Wyckoff sites 2a or 2b with symmetry 4/mmm, whereas for the materials $BaMn_2P_2$ and $BaMn_2As_2$ paramagnetic Mn ions use sites 4d that have a different symmetry, namely, $\bar{4}m2$. In the present case, our unit-cell structure factor for magnetic properties, equation (2), complies with the full magnetic space-group. It manifestly includes Mn site symmetry, because the expression in (2) is proportional to the Mn multipole that conforms to site symmetry $\bar{4}$m'2'.

## III. DIFFRACTION AMPLITUDES

We choose to describe electronic degrees of freedom of Mn ions in terms of (irreducible) spherical multipoles $\langle O^K_Q \rangle$, where $\langle ... \rangle$ denotes an expectation value (time-average) and the integer K is the multipole rank, with projections $-K \leq Q \leq K$. In the electronic structure factor,

$$\Psi^K_Q = \sum_\mathbf{d} \exp(i\mathbf{d} \cdot \mathbf{\tau}) \langle O^K_Q \rangle_\mathbf{d}, \qquad (1)$$

the sum is over sites labelled **d** in the magnetic unit-cell used by Mn ions, namely, Wyckoff 4d in I4'/m'm'm. The Bragg wavevector $\mathbf{\tau} = (h, k, l)$, and I-centring demands the condition $h + k + l$ even on the integer Miller indices. One finds,

$$\Psi^K_Q = 2 \langle O^K_Q \rangle \exp(i\pi l/2) (-1)^k [1 + (-1)^l \sigma_\theta \sigma_\pi], \qquad (2)$$

where $\sigma_\theta$ (time) and $\sigma_\pi$ (parity) are signatures = $\pm 1$ of the discrete symmetries of $\langle O^K_Q \rangle$. Evaluated for the paramagnetic region, with $\sigma_\theta = +1$, $\Psi^K_Q$ is identical to the result published previously [9]. The four sites in the unit cell that contribute in (2) are,

(1/2, 0, 1/4)↑, (1/2, 0, 3/4)↓, (0, 1/2, 3/4)↑, (0, 1/2, 1/4)↓,

and arrows indicate relative moment directions along the c-axis. Environments at the first pair of sites are related by the operation $2_y \equiv m'_z$, where Cartesian coordinates (x, y, z) coincide with cell edges in Fig. 1, and the second pair are displaced by (1/2, 1/2, 1/2). The results,

$$\langle O^K_{-Q} \rangle = (-1)^K \sigma_\theta \sigma_\pi \langle O^K_Q \rangle, \quad (-1)^p = \sigma_\pi, \text{ with } Q = \pm 2p, \qquad (3)$$

are required by site symmetry is $\bar{4}$m'2' [11], and they are used in the derivation of (2). Setting Miller indices equal to zero in (2) shows that, the unit-cell structure factor for bulk properties vanishes for conventional magnetic multipoles ($\sigma_\theta = -1$, $\sigma_\pi = +1$), as expected for compensated magnetic order. It is different from zero for Dirac multipoles that are both time-odd ($\sigma_\theta = -1$) and parity-odd ($\sigma_\pi = -1$), because the magnetic crystal-class 4'/m'm'm is one of the 58 crystal classes that allow the linear magnetoelectric effect. Dirac monopoles and dipoles (also called anapoles or toroidal dipoles) are forbidden, because |Q| = 2 is a minimum magnitude of projections when $\sigma_\pi = -1$. The electronic structure factor (2) is proportional to the Mn multipole, and this finding is explicit proof that discrete symmetries required by $\bar{4}$m'2' are necessary to get the complete description of the magnetic properties of BaMn$_2$P$_2$ and BaMn$_2$As$_2$, including their bulk properties. This is a general requirement, which becomes trivial in the singular case of identical site and magnetic crystal-class symmetries.

Amplitudes for x-ray or neutron diffraction derived from (2) contain Hermitian multipoles only, for which $\langle O^K_{-Q}\rangle = (-1)^Q \langle O^K_Q\rangle^*$ and identities,

$$\langle O^K_{-Q}\rangle = \langle O^K_Q\rangle^* = (-1)^K \sigma_\theta \sigma_\pi \langle O^K_Q\rangle, \qquad (4)$$

follow through correct use of (3). Parity-even multipoles, with $\sigma_\theta = -1$ & $\sigma_\pi = +1$, that deflect neutrons, $\langle T^K_Q\rangle$, are purely real (imaginary) for K odd (even), and the allowed projections include Q = 0, ± 4. A dipole $\langle T^1_0\rangle \equiv \langle T^1_z\rangle$ is proportional to the magnetic moment, in a first approximation.

## IV. NEUTRON DIFFRACTION

Neutron polarization analysis can be used to extract the magnetic contribution to the intensity of a Bragg spot with overlapping nuclear and magnetic amplitudes. Primary and secondary polarizations are **P** and **P'**, and a fraction $(1 - \mathbf{P} \cdot \mathbf{P'})/2$ of neutrons participate in events that change (flip) the neutron spin orientation. For a collinear magnetic motif one finds $(1 - \mathbf{P} \cdot \mathbf{P'})/2 \propto \{(1/2)(1 + P^2)|\langle \mathbf{Q}_\perp\rangle|^2 - |\mathbf{P} \cdot \langle \mathbf{Q}_\perp\rangle|^2\}$. A quantity SF = $|\langle \mathbf{Q}_\perp\rangle - \mathbf{P}(\mathbf{P} \cdot \langle \mathbf{Q}_\perp\rangle)|^2$ obtained with $P^2 = 1$ is a convenient measure of the strength of spin-flip scattering.

Bragg spots created by conventional magnetic multipoles are indexed by h + k odd and l odd, whereas Bragg spots created by Dirac multipoles are indexed by h + k even and l even. The magnetic neutron scattering amplitude $\langle \mathbf{Q}_\perp\rangle = [\boldsymbol{\kappa} \times (\langle \mathbf{Q}\rangle \times \boldsymbol{\kappa})]$ where $\boldsymbol{\kappa}$ is a unit vector in the direction of the Bragg wavevector, $\boldsymbol{\kappa} = \boldsymbol{\tau}(h, k, l)/|\boldsymbol{\tau}(h, k, l)|$. For the parity-even intermediate amplitude (l odd),

$$\langle \mathbf{Q}\rangle^{(+)} = 4\exp(i\pi l/2)(-1)^k \langle j_0(k)\rangle (g/2)(0, 0, \langle S_z\rangle); \text{ s-state ion}, \qquad (5)$$

which applies to $^6S$ for $Mn^{2+}$ ($3d^5$) with no orbital angular momentum $\langle \mathbf{L}\rangle = 0$ and a gyromagnetic factor g = 2.0. In this expression, k is the magnitude of $\boldsymbol{\tau}(h, k, l)$, and $\langle j_0(k)\rangle$ is a standard radial integral defined such that $\langle j_0(0)\rangle = 1$ [16]. However, $\langle \mathbf{Q}\rangle^{(+)}$ contains additional multipoles for all atomic configurations other than a pure s-state, and they are accompanied by radial integrals $\langle j_n(k)\rangle$ with n = 2, 4, ... that vanish at k = 0 [17, 18]. More generally, the so-called dipole-approximation,

$$\langle \mathbf{T}^1\rangle \approx (3/2)[2\langle \mathbf{S}\rangle \langle j_0(k)\rangle + \langle \mathbf{L}\rangle \{\langle j_0(k)\rangle + \langle j_2(k)\rangle\}], \qquad (6)$$

with $\langle j_2(0)\rangle = 0$ is useful and,

$$\langle \mathbf{Q}\rangle^{(+)} \approx (C/3)(0, 0, \langle T^1_z\rangle) \text{ with } C = 4\exp(i\pi l/2)(-1)^k. \qquad (7)$$

The factor C is purely imaginary in the present case, $h + k$ odd and $l$ odd. Electronic states with like parities make contributions to $\langle \mathbf{Q} \rangle^{(+)}$. Multipoles with an even rank, quadrupoles (K = 2) and hexadecapoles (K = 4), measure J-mixing in allowed states since even rank $\langle \mathbf{T}^K \rangle$ are zero within a J-manifold [17, 18].

Confining our immediate attention to multipoles up to and including octupoles allowed in the magnetic space-group I4'/m'm'm, parity-even magnetic amplitudes are ($l$ odd),

$$\langle \mathbf{Q}_x \rangle^{(+)} \approx - C\,(3/4)\,\sqrt{7}\,\kappa_x \kappa_z \langle T^3_0 \rangle, \quad \langle \mathbf{Q}_y \rangle^{(+)} \approx - C\,(3/4)\,\sqrt{7}\,\kappa_y \kappa_z \langle T^3_0 \rangle,$$

$$\langle \mathbf{Q}_z \rangle^{(+)} \approx C\,(3/2)\,[\langle T^1_z \rangle + (1/4)\,\sqrt{7}\,(3\kappa_z^2 - 1)\langle T^3_0 \rangle]. \qquad (8)$$

The octupole, $\langle T^3_0 \rangle$, is purely real and a linear combination of radial integrals $\langle j_2(k) \rangle$ & $\langle j_4(k) \rangle$. It is identically zero for a pure s-state ion. In the event that $\langle T^3_0 \rangle$ is different from zero, because the manganese ion configuration departs from a pure s-state, one or other of the contributions $\langle \mathbf{Q}_x \rangle^{(+)}$ & $\langle \mathbf{Q}_y \rangle^{(+)}$ will contribute at allowed Bragg spots since they are indexed by $\kappa_z \propto l$ odd, and $h + k$ odd. The allowed hexadecapole, $\langle T^4_{+4} \rangle$, is purely imaginary and has its origin in the orbital-spin part of the neutron electron interaction about which we have more to say in the context of Dirac multipoles.

Neutron diffraction by Dirac multipoles at sites occupied by Mn ions in $BaMn_2P_2$ and $BaMn_2As_2$ is determined by the amplitudes,

$$\langle \mathbf{Q}_x \rangle^{(-)} = C\,\kappa_x\,[Z_0 - Z_1\,(5\kappa_z^2 - 1) + Z_3\,(\kappa_x^2 - 3\kappa_y^2)],$$

$$\langle \mathbf{Q}_y \rangle^{(-)} = C\,\kappa_y\,[-Z_0 + Z_1\,(5\kappa_z^2 - 1) + Z_3\,(3\kappa_x^2 - \kappa_y^2)],$$

$$\langle \mathbf{Q}_z \rangle^{(-)} = C\,Z_2\,\kappa_z\,(\kappa_x^2 - \kappa_y^2), \qquad (9)$$

with C purely real for $l$ even ($h + k$ even). The quantity $Z_0$ contains only quadrupoles (K = 2), while $Z_1, Z_2, Z_3$, are linear combinations of quadrupoles, octupoles (K = 3) and hexadecapoles (K = 4). Bragg spots (0, 0, $l$) have no magnetic intensity, while $\langle \mathbf{Q}_x \rangle^{(-)} = - \langle \mathbf{Q}_y \rangle^{(-)}$ with $\langle \mathbf{Q}_z \rangle^{(-)} = 0$ for ($h, h, 0$) leads to $|\langle \mathbf{Q}_\perp \rangle^{(-)}|^2 = C^2\,[Z_0 + Z_1 - Z_3]^2$. Regarding spin-flip scattering, SF($h, h, 0$) = $|\langle \mathbf{Q}_\perp \rangle^{(-)}|^2$ for (a) neutron polarization $\mathbf{P}$ parallel to $\boldsymbol{\kappa}$ and (b) $\mathbf{P}$ normal to the plane of scattering ($\mathbf{P} \cdot \boldsymbol{\kappa} = 0$), whereas SF($h, h, 0$) = 0 for (c) $\mathbf{P}$ in the plane of scattering so that $\mathbf{P} \cdot \boldsymbol{\kappa} = 0$. Results quoted for the three values of the polarization, labelled (a), (b) & (c), illustrate a general condition $SF_a = SF_b + SF_c$.

Complete details of Dirac multipoles in neutron scattering, and the accompanying radial integrals, are found in references [18, 19]. By way of relevant examples,

$$\langle H^2_{+2}(1) \rangle' = (1/2)\,(h_1)\,\sqrt{(5)}\,\langle S_x\,n_x - S_y\,n_y \rangle,$$

$$\langle O^2_{+2}(1, a)\rangle' = -(1/4)\,(j_2)\,\sqrt{(15)}\,\langle L_x n_x - L_y n_y + \text{h.c.}\rangle, \tag{10}$$

$$\langle H^2_{+2}(3)\rangle' = (1/2)\,(h_3)\,\sqrt{(5/6)}\,\langle -2(S_x n_x - S_y n_y) + 5(n_x^2 - n_y^2)(\mathbf{S}\cdot\mathbf{n})\rangle,$$

$$\langle H^3_{+2}(3)\rangle'' = -(7/2)\,(h_3)\,\sqrt{(5/6)}\,\langle S_z n_z(n_x^2 - n_y^2) + \text{perm}\rangle$$

$$= (7/2)\,(h_3)\,\sqrt{(5/6)}\,\langle S_x n_x(1 - 3n_y^2) - S_y n_y(1 - 3n_x^2) - (n_x^2 - n_y^2)(\mathbf{S}\cdot\mathbf{n})\rangle,$$

$$\langle H^4_{+2}(3)\rangle' = (1/2)\,(h_3)\,\sqrt{(15/2)}\,\langle S_x n_x(3 - 7n_x^2) - S_y n_y(3 - 7n_y^2)$$

$$+ 3(n_x^2 - n_y^2)(\mathbf{S}\cdot\mathbf{n})\rangle, \tag{11}$$

occur in $Z_1$, $Z_2$ & $Z_3$. In these expressions for multipoles $\langle H^{K'}_{+2}(K)\rangle$, **n** is a unit vector for the position of an electron, and ' (") denotes the real (imaginary) part of a multipole. Irreducible multipoles created from a tensor product are discussed in the Appendix, with $\mathbf{H}^{K'}(K)$ tensor products of **S** and a normalized spherical harmonic $\mathbf{C}^K(\mathbf{n})$. In consequence, $K' = K - 1, K, K + 1$ with K odd in a parity-odd multipole. Careful calculations of Dirac multipoles for different materials are reported in several places that include references [13, 20, 21, 22].

The expectation value $\langle (n_x^2 - n_y^2) S_z n_z\rangle$ might be particularly large, because $S_z$ could be assigned a value close to 5/2 in a naive estimation, and it occurs in the quadrupole, octupole and hexadecapole found in (11). The expectation value in question occurs in the ratios 5: 7: − 9 in $\langle H^2_{+2}(3)\rangle'$: $\langle H^3_{+2}(3)\rangle''$: $\langle H^4_{+2}(3)\rangle'$.

Previously, $\langle (n_x^2 - n_y^2) S_z n_z\rangle$ was erroneously identified as a unique hexadecapole (K = 4) order parameter, whereas it is shown by us to contribute to irreducible multipoles of order K = 2 and K = 3 in addition [13]. Note that $\langle H^4_{+2}(3)\rangle' = (1/2)\,(h_3)\,\sqrt{(3/2)}\,\langle M^+_{42}\rangle$, where $M^+_{42}$ appears in the list of unique magnetic multipoles provided in Table II in reference [13]. The error by Watanabe and Yanase [13] is likely caused by omission of Mn site symmetry in the analysis.

Radial integrals $(j_n)$ and $(h_n)$ in (10) and (11) are formed with radial densities that belong to states with opposite parities, and some illustrative examples are found in references [18]. Substantial spin-dependent hybridization between d and pnictogen p states is reported from simulations of the electronic structure of $BaMn_2As_2$ - stronger hybridization than in the antimonide [23]. Similar behaviour is reported for the phosphor compound [24]. One finds $(j_2) \rightarrow (k\langle 3d|R|4p\rangle)/15$ for small values of $ka_o$, where $a_o$ is the Bohr radius. Contributions to the electric dipole moment $\langle 3d|R|4p\rangle$ are discussed in reference [19]. To the extent that radial wavefunctions are sensibly hydrogenic in form, ⟨3d|R|4p⟩ is proportional to $1/Z_c$ where $Z_c$ is the effective core charge seen by the electron. Explicit calculations using hydrogenic radial densities yield the results, $(h_1) \to 0.43\ (ka_o/Z_c)$, $(j_2) \to (1/5)\ (h_1)$, and $(h_3) \to 16.03\ (ka_o/Z_c)^3$ for small $ka_o$.

## V. X-RAY DIFFRACTION

Studies of various materials using resonant x-ray Bragg diffraction, accompanied by supporting calculations, are reported in references [8, 17, 25-29], among many others. The scattering amplitude derived from quantum-electrodynamics is developed in the small quantity $E/mc^2$ where E is the primary energy and $mc^2 = 0.511$ MeV [17]. At the second level of smallness in this quantity the amplitude contains resonant processes that may dominate all other contributions should E match an atomic resonance $\Delta$. Assuming also that virtual intermediate states are spherically symmetric, the scattering amplitude $\approx \{F_{\mu'\nu}/(E - \Delta + i\Gamma/2)\}$ in the region of the resonance, where $\Gamma$ is the total width of the resonance. Intensity of a Bragg spot is proportional to $|F_{\mu'\nu}/(E - \Delta + i\Gamma/2)|^2$. The numerator $F_{\mu'\nu}$ is a unit-cell structure factor for Bragg diffraction in the scattering channel with primary (secondary) polarization $\nu$ ($\mu'$). In keeping with convention, $\sigma$ denotes polarization normal to the plane of scattering, and $\pi$ denotes polarization within the plane of scattering. Our unit-cell structure factors include their dependence on a rotation of the crystal through an angle $\psi$ around the Bragg wavevector in a so-called azimuthal-angle scan. The Bragg angle is denoted by $\theta$ and the primary beam is deflected through an angle $2\theta$.

An electric dipole - electric quadrupole (E1-E2) event gives access to Dirac multipoles that have ranks K = 1, 2 & 3, and they have been labelled $\langle G^K_Q \rangle$ [17]. Since magnetic symmetry restricts projections Q to $\pm 2$, the anapole is forbidden. Bragg diffraction enhanced by the Mn L-edges at $\Delta \approx 0.65$ keV is not possible, since reflections are not within the Ewald sphere. In consequence, we consider enhancement at the Mn K-edge at $\Delta \approx 6.54$ keV, which equates to a photon wavelength $\approx 1.90$ Å. In so far as hydrogenic forms of radial wavefunctions are appropriate for the photo-ejected electron and empty valence states, radial matrix elements in an E1-E2 event are $\langle 1s|R|4p \rangle = 0.30\ (a_o/Z_c)$ and $\langle 1s|R^2|3d \rangle = 1.73\ (a_o/Z_c)^2$. Interference between parity-even (E2-E2) and parity-odd (E1-E2) diffraction amplitudes reported for corundum-like materials ($V_2O_3$ & $\alpha$-$Fe_2O_3$) will not occur in diffraction by the materials of interest here, because our two amplitudes are indexed by different Miller indices, namely, $l$ odd (E2-E2) and $l$ even (E1-E2) [8].

K-edge amplitudes for Bragg diffraction are composed of $\langle \mathbf{G}^2 \rangle \propto \langle \{\mathbf{L} \otimes \mathbf{n}\}^2 \rangle$ and $\langle \mathbf{G}^3 \rangle \propto \langle \{\{\mathbf{L} \otimes \mathbf{L}\}^2 \otimes \mathbf{\Omega}\}^3 \rangle$, where the orbital anapole $\mathbf{\Omega} = i[\mathbf{L}^2, \mathbf{n}] = (\mathbf{L} \times \mathbf{n} - \mathbf{n} \times \mathbf{L})$. By way of illustrating the information content of E1-E2 unit-cell structure factors, let us consider Bragg spots indexed by Miller indices $l = 0$ and $h = k$. Structure factors are found to be,

$$F_{\sigma'\sigma}(h, h, 0) = (C/\sqrt{15})\cos(\theta)\sin(\psi)\ [\sqrt{2}\ \langle G^2_{+2} \rangle' + (3\cos(2\psi) + 1)\ \langle G^3_{+2} \rangle''],$$

$$F_{\pi'\sigma}(h, h, 0) = (C/2\sqrt{15}) \sin(2\theta) \cos(\psi) [2\sqrt{2} \langle G^2_{+2}\rangle'$$

$$+ (3 \cos(2\psi) - 1) \langle G^3_{+2}\rangle"]. \quad (12)$$

Enhancement at the Mn K-edge gives access to reflections with $h$ = 1, 2 & 3. Azimuthal-angle scans in the unrotated and rotated channels of polarization evidently provide excellent opportunities to determine a relative strength for the Dirac quadrupole and octupole. The rotated channel does not contain Thompson scattering, which is an advantage when contributions from weak Dirac multipoles are of interest. To illustrate the pronounced sensitivity to the ratio of the two multipoles of intensity in a Bragg spot observed within the rotated channel of polarization, Fig. 2 displays a reduced version of $|F_{\pi'\sigma}|^2$ as a function of the azimuthal angle for p = $(\langle G^3_{+2}\rangle"/(\langle G^2_{+2}\rangle' 2\sqrt{2}))$ = 0.5 & − 0.5. By contrast, Bragg spots indexed by (0, 0, $l$) are much less useful, because the the two multipoles in $F_{\pi'\sigma}$ possess identical azimuthal angle dependences. One finds,

$$F_{\pi'\sigma}(0, 0, l) = (C/\sqrt{30}) \cos^2(\theta) \cos(2\psi) [\langle G^2_{+2}\rangle' - 4\sqrt{2} \langle G^3_{+2}\rangle"], \quad (13)$$

with $l$ an even integer. At the origin of the azimuthal-angle scan, $\psi$ = 0, the crystal a-axis is normal to the plane of scattering. The two-fold symmetry of $F_{\pi'\sigma}(0, 0, l)$ with respect to $\psi$ is expected from the symmetry of a Mn environment.

## VI. DISCUSSION

Magnetoelectric properties of the layered, room-temperature antiferromagnets $BaMn_2P_2$ and $BaMn_2As_2$ are discussed making full use of information in the relevant magnetic space-group. The space-group I4'/m'm'm (#139.536 [11]), with Mn ions using Wyckoff positions 4d, is derived from knowledge of the chemical and magnetic structures reported in accounts of previous diffraction experiments [1, 2]. We go on to predict effects in neutron and resonant x-ray Bragg diffraction patterns that are unique signatures of a magnetoelectric state that supports Mn Dirac multipoles, which are time-odd (magnetic) and parity-odd (acentric). It is shown that the pattern of Bragg spots caused by Dirac multipoles and the pattern caused by a G-type antiferromagnetic order of conventional magnetic dipoles (time-odd and parity-even) are distinguishable, with Miller index $l$ even in one case and odd in the other. An improper rotation in Mn site symmetry, the unique roto-inversion $\bar{4}$, lies at the heart of the distinction in diffraction patterns.

For x-ray resonant Bragg diffraction, we predict great sensitivity in azimuthal-angle scans, where the crystal is rotated about the Bragg wavevector, to the relative size of Dirac multipoles. Structure factors for neutron diffraction that we report are exact, and contain all

Dirac multipoles allowed by symmetry. In turn, we are in a position to give a complete study of spin-flip scattering obtained from neutron polarization analysis [30].

A Dirac multipole calculated by Watanabe and Yanase [13] is relevant to the feasibility of diffraction by $BaMn_2As_2$, with grounds for optimism about the likely success of diffraction experiments to advance our knowledge about magnetoelectric properties of $BaMn_2As_2$ and $BaMn_2P_2$. However, our principal predictions, about diffraction by Dirac multipoles, cannot be derived from the theoretical framework employed by Watanabe and Yanase [13], because it omits all reference to Mn site symmetry. The fatal shortcoming in the framework is compounded by an inadequate account of primary and secondary order-parameters.

**ACKNOWLEDGEMENT** One of us (SWL) is grateful to Professor S. P. Collins (Diamond Light Source, Ltd.) for a discussion about the feasibility of the proposed x-ray diffraction experiments.

## APPENDIX

A normalized spherical harmonic $C^a{}_\alpha(\mathbf{n})$ is defined in the notation adopted by Racah, namely [18],

$$C^a{}_\alpha(\mathbf{n}) = [(4\pi)/(2a + 1)]^{1/2}\, Y^a{}_\alpha(\mathbf{n}), \tag{A.1}$$

where $\mathbf{n}$ is a unit vector, $a$ is the rank and projections $\alpha$ obey $-a \leq \alpha \leq a$. The complex conjugate satisfies $[C^a{}_\alpha(\mathbf{n})]^* = (-1)^\alpha C^a{}_{-\alpha}(\mathbf{n})$. A similar relation holds for Hermitian multipoles with $\langle O^K{}_Q \rangle = (-1)^Q \langle O^K{}_{-Q} \rangle^*$.

A spherical tensor of rank $K$ is constructed from the product of tensors $A^a$ and $B^b$ using,
$$\{A^a \otimes B^b\}^K{}_Q = \sum_{\alpha\beta} A^a{}_\alpha B^b{}_\beta\, (a\alpha b\beta | KQ)$$

$$= (1/2)\sum_{\alpha,\beta} \{[A^a{}_\alpha B^b{}_\beta + B^b{}_\beta A^a{}_\alpha] + [A^a{}_\alpha B^b{}_\beta - B^b{}_\beta A^a{}_\alpha]\}\, (a\alpha b\beta | KQ). \tag{A.2}$$

In the second equality the product of tensors is a sum of Hermitian and anti-Hermitian operators. The latter is zero if $A^a$ and $B^b$ commute, but $\{A^a \otimes B^b\}^K{}_Q$ is not Hermitian even in this case. The Clebsch-Gordan coefficient in (A.2) and Wigner 3-j symbol are purely real quantities related by [31],

$$(a\alpha b\beta | KQ) = (-1)^{-a+b-Q} \sqrt{(2K+1)} \begin{pmatrix} a & b & K \\ \alpha & \beta & -Q \end{pmatrix}. \tag{A.3}$$

The tensor product of entangled spin and orbital degrees of freedom used in Section IV of the main text is $\langle \mathbf{H}^K(K) \rangle \propto \{\mathbf{S} \otimes C^K(\mathbf{n})\}^{K'}$ with a proportionality factor chosen to make $\mathbf{H}^K(K)$ Hermitian.

A matrix element of a spherical tensor-operator obeys the Wigner-Eckart Theorem [31]. Denoting such an operator by $O^K_Q$,

$$\langle JMsl| O^K_Q |J'M's'l'\rangle = (-1)^{J-M} (Jsl\|O^K\|J's'l') \begin{pmatrix} J & K & J' \\ -M & Q & M' \end{pmatrix}, \quad (A.4)$$

in which $(Jsl\|O^K\|J's'l')$ is a so-called reduced matrix-element (RME), and total angular momentum $\mathbf{J} = \mathbf{s} + \mathbf{l}$. Note that $(J - M)$ is always an integer. Should $O^K$ be a function of orbital operators alone,

$$\langle lm| O^K_Q |l'm'\rangle = (-1)^{l-m} (l\|O^K\|l') \begin{pmatrix} l & K & l' \\ -m & Q & m' \end{pmatrix}. \quad (A.5)$$

An RME of an operator with defined discrete symmetries obeys two fundamental identities that apply for both $J$ integer and $J$ half-integer states [32],

$$(Jsl\|O^K\|J's'l') = (-1)^{J'-J} (J's'l'\|O^K\|Jsl)^*, \quad (A.6a)$$

$$(J's'l'\|O^K\|Jsl) = (-1)^{J-J'} \sigma_\theta \sigma_\pi (-1)^K (Jsl\|O^K\|J's'l'). \quad (A.6b)$$

The first identity holds for an Hermitian operator, while the second identity is independent of the specific operator, because it depends solely on definitions of time-reversed states and parity [17, 32]. In (A.6b) $\sigma_\theta = \pm 1$ ($\sigma_\pi = \pm 1$) is the time-signature (parity-signature) of $O^K$. In most cases of interest, a RME is either purely real or purely imaginary, in which case (A.6) tells us that $[\sigma_\theta \sigma_\pi (-1)^K] = \pm 1$, where the upper sign applies to purely real and the lower sign to a purely imaginary RME of an Hermitian operator.

The RME of a tensor product (A.2) formed by spin and spatial variables $z^a$ and $y^b$, respectively, is usefully written in terms of a unit tensor [17, 32]. We define such an RME as,

$$(\theta\|\{z^a \otimes y^b\}^K\|\theta') = (s\|z^a\|s) (l\|y^b\|l') W^{(a,b)K}(\theta, \theta'), \quad (A.7)$$

where $W^{(a,b)K}(\theta, \theta')$ is the unit tensor and composite labels $\theta = Jsl$ and $\theta' = J's'l'$. A tensor product is generally not an Hermitian operator, even when both parent operators, $z^a$ and $y^b$, are Hermitian. An Hermitian operator can be constructed from a tensor product, however, with a little ingenuity. Introduction of a complex phase factor suffices for commuting operators. In the particular case of one electron [32],

$$W^{(a,b)K}(\theta, \theta') = [(2j+1)(2K+1)(2j'+1)]^{1/2} \begin{Bmatrix} s & s & a \\ l & l' & b \\ j & j' & K \end{Bmatrix}, \quad (A.8)$$

in which $s = 1/2$ yields $a = 0$ or 1, with $(s\|z^0\|s) = \sqrt{2}$ or $(s\|z^1\|s) = \sqrt{(3/2)}$ in (A.7). The magnitude of the 9j-symbol in (A.8) is unchanged by an even or odd exchange of columns or rows, while its sign is changed by a factor $(-1)^\Re$ with an odd exchange of columns or rows where $\Re = (1 + a + l + l' + b + j + j' + K)$ [30]. A single operator in a coupled scheme, using spin and orbital quantum labels, possesses a RME proportional to a 6j-symbol [31]. We use S-L coupling, for which the RME of an operator $O^K$ working only on part 2 (orbital variables) is,

$$(J'sl'\|O^K\|Jsl) = (-1)^{s+l+J'+K}[(2J+1)(2J'+1)]^{1/2}(l'\|O^K\|l)\begin{Bmatrix} l' & J' & s \\ J & l & K \end{Bmatrix}. \qquad (A.9)$$

The result (A.9) provides $(J'sl'\|\{L \otimes n\}^K\|Jsl)$ in terms of $(l'\|\{L \otimes n\}^K\|l)$, which can be obtained by straightforward use of (A.5) for the tensor product derived from (A.2).

**Table I**. ThCr$_2$Si$_2$-type magnetoelectric materials BaMn$_2$P$_2$ [1] and BaMn$_2$As$_2$ [2]. Data gathered on powder samples using neutron Bragg diffraction. Space group I4/mmm (#139, tetragonal 4/mmm crystals class, Miller indices $h + k + l$ even from I-centring) with Mn in 4d, Ba in 2a and P, As in 4e. Magnetic space-group I4'/m'm'm (#139.536, 4'/m'm'm magnetoelectric crystal-class) [11], with Mn dipole moments parallel to the c-axis in a collinear G-type antiferromagnetic motif - magnetic propagation vector **k** = (0, 0, 0) - and indexed by Miller indices $h + k$ odd and $l$ odd. Saturation magnetic moment, μ, is less than 5 μ$_B$ expected of the high-spin Mn$^{2+}$ (3d$^5$) configuration. A continuous magnetic phase transition, not exhibiting magnetic hysteresis, with no change in the structure or distortion of the lattice is inferred from experimental evidence. Approximate cell parameters a & c, and Néel temperature T$_N$.

| Compound | a (Å) | c (Å) | μ (μ$_B$) | T$_N$ (K) |
| --- | --- | --- | --- | --- |
| BaMn$_2$P$_2$ [1] | 4.04 | 13.05 | 4.2 (at T=293K) | >750 |
| BaMn$_2$As$_2$ [2] | 4.15 | 13.41 | 3.88 (at T=10 K) | 625 |

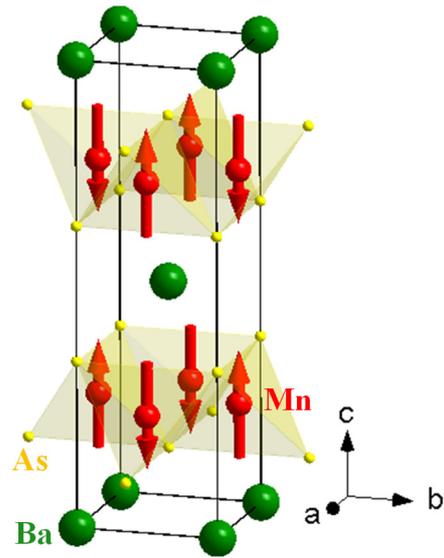

Fig. 1. Chemical (I4/mmm) and magnetic (I4'/m'm'm) structures of BaMn$_2$As$_2$ [2]. Red arrows denote Mn dipole moments parallel to the c-axis in a G-type antiferromagnetic motif, with Ba ions (green) and As ions (yellow). Cartesian coordinates (x, y, z) used in the text coincide with cell edges.

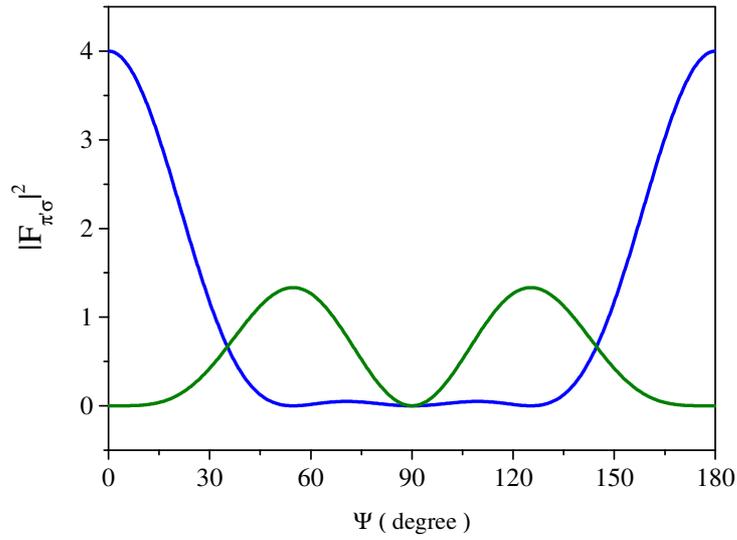

Fig. 2. Bragg spots ($h$, $h$, 0) at which diffraction by conventional (parity-even) magnetism is forbidden. Intensity in the rotated channel of polarization $|F_{\pi'\sigma}|^2$ is illustrated as a function of the azimuthal angle $\psi$, using an expression derived from (12). The c-axis is normal to the plane of scattering and parallel to $\sigma$ polarization. The quantity $p = (\langle G^3_{+2}\rangle''/(\langle G^2_{+2}\rangle' \, 2\sqrt{2}))$ takes two values: $p = 0.50$ (blue line) and $p = -0.50$ (green line). The quantity plotted is $\{\cos(\psi) [1 + (3 \cos(2\psi) - 1) p]\}^2$ with $\psi$ in the range 0 - 180°.